\documentclass[12pt]{article}

\usepackage{bm}
\usepackage{amssymb}
\usepackage{enumerate}

\usepackage{graphicx} 

\begin{document}

\title{Light scalars, ($g_{\mu}-2$) muon anomaly and dark matter in a  model with a Higgs democracy}

\author{       N.V.~Krasnikov$^{1,2}$  
\\
$^{1}$ INR RAS, 117312 Moscow 
\\
$^{2}$ Joint Institute for Nuclear Research,141980 Dubna}




\date{\today}

\maketitle

\begin{abstract}

We consider  isosinglet scalar  extension of 
a model with a Higgs democracy - multihiggs extension of the SM where each quark and lepton 
 has its own Higgs isodoublet. 
The addition of light isosinglet scalar 
 allows to solve both muon $g_{\mu}-2$ anomaly and dark matter problem.   
Also we point out that an  extension of the model with $L_{\mu} - L_{\tau}$ vector interaction 
allows not only explain    muon $g_{\mu} - 2$ anomaly but also dark matter density.

\end{abstract}

keywords:    {light dark matter, muon anomaly, light vector and scalar messengers}

\newpage

\section{Introduction}

At present there are several   signals that    new physics beyond the SM exists. The most solid fact is the existence of
dark matter  \cite{Universe1} -\cite{Universe3}. The nature of dark matter is one of challenging questions in modern physics. 
There are a lot of candidates on the role of dark matter  \cite{Universe1} - \cite{Universe3}.   
In particular, models with light vector(scalar) bosons 
with a mass $m_d \leq O(10)~GeV$ \cite{lightdark,lightdark1}  are  rather popular now. 
The main idea  is that new light vector(scalar) bosons    are   mediators 
connecting   our world and the world of dark matter particles\cite{lightdark1}. 
Other possible hint in favour of new physics beyond the SM is 
muon $g_{\mu}-2$ anomaly, namely the precise measurement of the anomalous magnetic
moment of the positive muon from the
Brookhaven AGS experiment \cite{g-2} gives a result which
is $3.6 \sigma$ higher than the Standard Model (SM) prediction\footnote{Here $a_{\mu} \equiv \frac{g_{\mu} -2}{2}$.} 
\begin{equation}
a_{\mu}^{exp} - a_{\mu}^{SM} = (288 \pm 80) \cdot 10^{-11} \,.
\end{equation} 
There are a lot of  $g_{\mu}-2$ anomaly explanations.
In particular, one of the possible    
explanations assumes the existence of new relatively light(with 
a mass $m_{Z^`} \leq 1~GeV$) vector boson(dark photon) wich couples very weakly with muon with the 
coupling constant $\alpha^`_{Z^`} \sim O(10^{-8})$ \cite{vecmuon1}- \cite{vecmuon8}. Recent experimental data 
\cite{NA64}-\cite{BABAR1} severely restrict\footnote{The model with light 
vector boson interacting with the SM electromagnetic current is excluded as an explanation of 
$g_{\mu}-2$ anomaly. However, the model with vector boson interacting with $L_{\mu} - L_{\tau}$ current 
survives for $m_{Z^`} \leq 210~MeV$ \cite{vecmuon8}. Also more exotic model with infinite number of 
light vector bosons\cite{vecmuon8} survives.}  
 but not completely exclude  this possibility of the $g_{\mu}-2$ anomaly explanation. Recent claim \cite{17 MeV} 
of the discovery of  $~17~MeV$ vector particle observed 
as a peak in $e^+e^-$ invariant mass distribution in nuclear transitions makes the question of possible light 
vector boson    existence  
extremely interesting and enhance motivation for further experimental and theoretical studies. 
Other related explanation of $g_{\mu}-2$  assumes  the existence 
of new relatively light scalar particle \cite{Mi1} - \cite{POSPELOV1}. 
However, the simplest extension of the SM with additional scalar field $\phi$ can't explain 
      $g_{\mu}-2 $ anomaly and(or) current dark matter density. The reason is that in the SM 
the Yukawa coupling constant of the muon with the SM Higgs isodoublet 
$h_{\mu} = \frac{m_{\mu}}{<H>} \approx 6 \cdot10^{-4}$
is extermely small. The Yukawa coupling $g_{\mu}$ of muons with the scalar $\phi$ is 
$g_{\mu} =      h_{\mu}  \sin \theta_{h \phi}$, where $ \theta_{h \phi}$   is the mixing angle of 
the scalar $\phi$ with the Higgs field\footnote{In the unitaire gauge $H = (0,~\frac{h}{\sqrt 2} + <H>)$, $<H> = 174~GeV$.}  $h$. 
Experimental bounds from $K$- and $B$- meson decays lead to upper bound $|\sin \theta_{h \phi}| \leq O(10^{-3})$ \cite{limit,limit1} 
that excludes this model.   

In this paper we consider  isosinglet scalar  extension of 
the model with a Higgs democracy  \cite{krasnikov1992} - multihiggs extension of the SM where each quark and lepton 
has its own Higgs isodoublet. 
 The existence of isosinglet scalar field allows to solve  $g_{\mu}-2$ anomaly and it  also  serves as a messenger between the SM 
matter and  dark matter \cite{lightdark,lightdark1}\footnote{In ref.\cite{POSPELOV1} the SM extension with additional 
scalar field and
Higgs isodoublet $H_{ad}$ 
which couples only with leptons  has been proposed. The model \cite{POSPELOV1} 
can explain the $g_{\mu} -2$ anomaly.}.  Due to small mixing of isosinglet  scalar with the Higgs boson responsible for 
top quark mass the proposed model escapes bounds from rare $K$- and  $B$-meson decays 
and solves  both  $g_{\mu}-2$ and dark matter problems.  
Also we point out that an  extension of the model with $L_{\mu} - L_{\tau}$ vector interaction  
of new light vector boson 
allows not only explain     $g_{\mu}-2$ anomaly but also dark matter density.  

This paper is organized as follows. In the next section we describe model with three lepton Higgs isodoublets,  scalar isosinglet and light dark matter. 
We show that proposed model is able to explain simultaneously both muon $g_{\mu} - 2$ anomaly and today dark matter density. In 
section 3 we point out that dark matter extension of the model with light vector boson interacting with 
$L_{\mu} - L_{\tau}$ current can explain not only $g_{\mu} -2$ anomaly but also today dark matter density. We present our 
conclusions in section 4. The appendices A and B collect main formulae for new physics $g_{\mu} -2$ contributions 
and dark matter annihilation cross sections correspondingly.

\section{Model with three lepton Higgs isodoublets,  scalar isosinglet and light dark matter}

In this paper we are  interested mainly in leptonic sector so    we  consider 
 the particular  case where all quark masses  arise due to nonzero 
vacuum expectation value $<H> $ of the Higgs isodoublet $H$. 
In our model each lepton generation $l = e, \mu, \tau$ has its own Higgs isodoublet $H_l$.
Besides we introduce additional scalar field $\phi$ and Dirac fermion dark matter field $\psi_d$. 
The Lagrangian of the model has the form
\begin{equation}
L_{tot} = L_{SM, h_l = 0} + L_{Yuk,H_l}  + L_{H_l} +  L_{\phi \psi} +  L_{HH_l\phi} \,,
\end{equation}
where $L_{SM,h_l=0}$ is the SM Lagrangian with all lepton Yukawa couplings equal  to zero and
\begin{equation}
L_{Yuk,H_l} = -\sum_{l}h_l\bar{A}_ll_RH_l  + H.c.\,,
\end{equation}
\begin{equation}
L_{H_l}  = \sum_{l,l^`}[\Delta^{\mu} H^+_l \Delta_{\mu}H_l   -   m^{2}_{h_l}H^+_lH_l - \lambda_{ll^`}(H^+_lH_l)(H^+_{l^`}H_{l^`})] \,,
\end{equation}
\begin{equation}
L_{\phi\psi} = \frac{1}{2} \partial^{\mu}\phi\partial_{\mu}\phi 
 -\lambda_{\phi}(\phi^2 - c^2)^2 +i\bar{\psi}_d\hat{\partial}\psi_d - g_{\psi}\phi \bar{\psi}_d\psi_d  \,,
\end{equation}
\begin{equation}
L_{HH_l\phi} = \sum_{l}\sqrt{2}M_lH_l^+H\phi + H.c. \,.
\end{equation}
Here 
$$
(\nu, e)_L = A_e, ~(\nu^`, \mu)_L = A_{\mu}, ~(\nu^{``}, \tau)_L = A_{\tau} \,,
$$
$$
 l_R = e_R, \mu_R, \tau_R 
$$
and $H_{e} = (H^+_{e},~H^0_{e})$, $H_{\mu} = (H^+_{\mu},~H^0_{\mu})$, $H_{\tau} = (H^+_{\tau},~H^0_{\tau})$ 
are  Higgs isodoublets responsible for $e$-, ~$\mu$- and $\tau$- lepton masses. 
The Lagrangian (2) is invariant under the symmetries
\begin{equation}
A_l \rightarrow - A_l \,, ~~l_R \rightarrow - l_R  \,,
\end{equation} 
\begin{equation}
H_l \rightarrow -H_l, ~~\phi \rightarrow - \phi \,, ~~l_R \rightarrow - l_R , ~~\psi_d \rightarrow -\psi_d \,,
\end{equation} 
The interaction  $L_{HH_l\phi}$ is superrenormalizable, it does not 
generates new types of ultraviolet divergences and it is the single interaction 
connecting dark matter particles $\psi_d$ with our world.
The discrete symmetry (8) prohibits  mixing terms like $H^+_lH, ~~\phi H^+H, ~~\phi H^+_lH_l$.  
Before the discrete symmetry breaking (8) 
the vacuum expectation values of the lepton isodoublets are equal to zero. 
After the symmetry breaking $<\phi> \neq 0$ and nonzero  mixing terms 
$\sqrt{2}<\phi> M_lH^+H_l$ are generated that leads to nonzero $<H_l> $\footnote{$H_l = (H^+_l, \frac{1}{\sqrt 2}(h_{ls} + ia_l) + <H_l>).$} 
\begin{equation}
<H_l>  = \sqrt{2}\frac{M_l<H><\phi>}{m^2_{h_l}}  \,
\end{equation}
and hence to nonzero lepton masses. 
As a result of the symmetry breaking the dark matter fermion 
$\psi_d$ acquires a mass $m_d = g_{\psi}<\phi>$.     
Trilinear term (6) leads to the mixing of the scalar field $\phi^` \equiv \phi - <\phi> $ with 
leptonic Higgs scalars $h_{ls}$.  As a consequence of   $<H> \neq 0$ the 
mixings between scalar fields $h_{ls}$ and $\phi$ are  generated. For small $M_l<\phi> ~\ll ~m^2_{h_l}$ and 
$m^2_{\phi} ~\ll ~m^2_{h_l}$
the mixings are 
\begin{equation}
\theta_{hh_l} = -\frac{\sqrt{2} M_l<\phi>}{m^2_{h_l}} \,,
\end{equation}
\begin{equation}
\theta_{h\phi} = -\frac{\sqrt{2} M_l <H_l>}{m^2_{h}} \,,
\end{equation}
\begin{equation}
\theta_{h_l\phi} = -\frac{\sqrt{2} M_l <H>}{ m^2_{h_l}} \,,
\end{equation}
As a consequnce of the relations (11,12) we find that 
\begin{equation}
\theta_{h\phi} = \frac{<H_l>}{H} \frac{m^2_{h_l}}{m^2_{h}}\theta_{h_l\phi} \,.
\end{equation}
Nonzero  ($\phi$, $h_l$) mixing leads to  induced Yukawa interaction 
\begin{equation}
L_{ll\phi^`_k} =  \sum_{l}g_l  \phi^` \bar{l} l \,,
\end{equation}
where
\begin{equation}
g_l  = h_l\sin \theta_{h_l\phi} \,.
\end{equation}
Note that there are strong bounds on mixing angle $\theta_{h \phi}$  \cite{limit, limit1}
derived from data on rare $K$-,  $B$-meson decays and 
invisible Higgs boson decay $h \rightarrow invisible $.  
The most stringent bound 
\begin{equation}
|\theta_{h \phi}| \leq 1.6 \cdot 10^{-4}
\end{equation} 
was obtained from the experimentally measured decay width of the $K^+ \rightarrow \pi^+ \nu\bar{\nu}$ decay 
and it is valid for $m_{\phi} < m_{K^+} - m_{\phi}$.

\subsection{Muon anomalous magnetic moment}

The precise measurement of the anomalous magnetic
moment of the positive muon from the
Brookhaven AGS experiment \cite{g-2} gives a result which
is $3.6 \sigma$ higher than the Standard Model (SM) prediction 
\begin{equation}
a_{\mu}^{exp} - a_{\mu}^{SM} = (288 \pm 80) \cdot 10^{-11} \,,
\end{equation} 
where $a_{\mu} \equiv \frac{g_{\mu} -2}{2}$.
Additional contribution (33) allows to expain the Brookhaven AGS result. 
For instance, for $ m_{\phi} = 0.001~GeV$,   $m_{\phi} = 0.1~GeV$, $m_{\phi} = 1~GeV$, $m_{\phi} = 5~GeV$ 
and   $m_{\phi} = 10~GeV$ 
the coupling constants $\alpha_{s\mu} \equiv \frac{g^2_{\mu}}{4\pi} = (1.2 \pm 0.34)\cdot 10^{-8}$, 
 $\alpha_{s\mu}= (3.5 \pm 1.0)\cdot 10^{-8}$,
 $\alpha_{s\mu}= (4.7 \pm 1.3)\cdot 10^{-7}$, 
$\alpha_{s\mu}= (6.2 \pm 1.7)\cdot 10^{-6}$, 
and  $\alpha_{s\mu}= (2.0 \pm 0.57)\cdot 10^{-5}$
reproduce   $g_{\mu}-2_{\mu}$ anomaly (1).

One can show that the bound on mixing angele $\theta_{h \phi}$ does not contradict to the values 
of the mixing angle $\theta_{h_{\mu} \phi}$ explaining muon $(g-2)_{\mu}$ anomaly. For instance, 
for Yukawa coupling $h_{\mu} = 1$,  $m_{h_{\mu}} = 750~GeV$ and $m_{\phi} = 100~MeV$  
the mixing angle $\theta_{ h_{\mu} \phi} = 6.6 \cdot 10^{-4}$  explains the $g_{\mu}-2$ anomaly.
As a consequence of the relation (13) the 
   mixing angle   $\theta_{h \phi} =  1.4 \cdot  10^{-5}$ that does not contradict to 
the experimental bound (16).


\subsection{Dark matter}

In propoded model fermion $\psi_d$ is a dark matter field and 
the scalar field $\phi$ plays a role of messenger.
To estimate the dark matter density we 
assume  that in the hot early Universe dark matter is in equilibrium with 
ordinary matter. During the Universe expansion the temperature decreases and at some 
temperature $T_d$ the thermal decoupling of the dark matter starts to work \cite{Universe1} - \cite{Universe3}. Namely, at 
some freeze-out temperature the annihilation cross-section  $DMparticles  ~~\rightarrow 
~~SM particles $
becomes too small to obey the equilibrium of dark matter particles with 
the SM particles and dark matter decouples.
In the  model with dark fermion and scalar mediator we have $p$-wave annihilation 
cross-section\footnote{CMB bound \cite{Planck} excludes 
s-wave annihilation for $m_d \leq 10~GeV$.}. 
The requirement that today dark matter density is $ \Omega_{d} h^2 \approx  0.12$
leads to the $p$-wave  cross section  
$ <\sigma v_{rel}><v^2_{rel}> =  2.6 \cdot 10^{-8}GeV^{-2} $(see formula (42) of the Appendix B).
The annihilation cross-section  
 $\sigma_{an}(\psi_d \bar{\psi}_d \rightarrow f \bar{f}) $ ($ f = e,\mu )$ for $m_d \gg m_f$ and $s \approx 4m^2_d$   is 
\begin{equation}
\sigma_{an}(\psi_d \bar{\psi}_d \rightarrow f \bar{f})v_{rel} = 
    \frac{g^2_dg^2_f m^2_d}{8 \pi (m^2_{\phi} - 4 m^2_{d})^2} v^2_{rel} \,,
\end{equation}
where $ g_f = g_{e}, g_{\mu} $ and $m_d$ is the mass of the dark matter fermion $\psi_d$. 
As a consequence of (18,42) we find that\footnote{Here $\alpha_d = \frac{g_{\psi}^2}{4\pi}$ and $\alpha_{s\mu} = \frac{g_{\mu}^2}{4\pi}$}
\begin{equation}
\alpha_d \alpha_{s\mu} =  0.41\cdot  10^{-8} \cdot (\frac{m_d}{GeV})^2\cdot (\frac{m^2_{\phi}}{m^2_{d}} - 4)^2 \,.
\end{equation}  
The requirement that one loop  $\phi$-boson contribution explains muon $g_{\mu}-2$ anomaly (1) 
allows to predict the dependence of the Yukawa coupling constant 
$\alpha_{s\mu}$  on the scalar boson mass $m_{\phi}$. 
Besides if we  require that the process of dark matter annihilation into $\mu^-\mu^+$ 
pair dominates we can use the formulae (18,42) to predict the dependence of the dark matter 
coupling constant $\alpha_{d} $ on the $\phi$-boson mass.   
As a numerical  example take   $m_{\phi} = 10~GeV[1~GeV]$ and  $m_d = 4~GeV[0.4~GeV]$.  
Using the formulae (18,33,42) one can find  that 
 $\alpha_{s \mu } =  (2.0  \pm 0.57) \cdot10^{-4}[ (4.7 \pm 1.3)\cdot 10^{-7}]$
and $ \alpha_{d} =   (0.017 \pm 0.05)[0.073 \pm 0.020] $
reproduce both the  $(g_{\mu}-2)$ anomaly  and today dark matter density. 

Note that in  our model the scalar messenger field $\phi$ interacts also with electron and $\tau$-lepton so 
 we can use other annihilation channels 
\begin{equation} 
\psi_d \bar{\psi}_d \rightarrow  e^+e^- \,,
\end{equation}
\begin{equation} 
\psi_d \bar{\psi}_d \rightarrow  \tau^+\tau^- \,
\end{equation}
for today dark matter explanation.

For $m_d < m_{\mu}$ the annihilation channel   $\psi_d \bar{\psi}_d \rightarrow \mu^+ \mu^- $ 
does not work and   the annihilation
 $\psi_d \bar{\psi}_d \rightarrow  e^+e^- $  into  
electron-positron pair can help. 
For electrons for $m_d \gg m_e$ 
and   $m_{\phi} = 2.5m_d$ we find
\begin{equation}
\alpha_{e \phi}\alpha_{d} =  2.1  \cdot 10^{-8}\cdot (\frac{m_d}{1~GeV})^2 \,.
\end{equation}  
As a consequence of the tree level unitarity  $ \alpha_d \leq 1 $ 
we obtain the lower bound 
\begin{equation}
\alpha_{e \phi} \geq   2.1  \cdot 10^{-8}\cdot (\frac{m_d}{1~GeV})^2 
\end{equation}  
on Yukawa coupling constant  $g^2_{e \phi}$ .


\section{Dark matter in a  model with $L_{\mu} - L_{\tau}$ current} 

The model with $L_{\mu} - L_{\tau}$ vector current   
  interaction  $L_{Z^`} = e_{\mu}Z^`_{\nu}[\bar{\mu}\gamma^{\nu}\mu - \bar{\tau}\gamma^{\nu}\tau]$ 
of new  abelian vector field $Z^{`}_{\mu}$ \cite{volkas}  with $\mu$ and $\tau$ 
leptons  allows to explain  $g_{\mu} - 2$ anomaly \cite{vecmuon1} - \cite{vecmuon8} 
and it does not contradict to existing  experimental bounds for $m_{Z^`} \leq 2 m_{\mu}$ \cite{vecmuon8}. 
Here we would like to mention that it is possible to construct
dark matter  extension of this model 
which explains today dark matter density. The simplest possibility is to add 
the complex scalar field\footnote{The annihilation cross-section for scalar dark matter 
has $p$-wave suppression that allows to escape CMB bound \cite{Planck}.} $\phi_d$.  The charged dark 
matter field  $\phi_d$ interaction with   the $Z^`_{\mu}$ field is 
$L_{\phi Z^`} = (\partial^{\mu}\phi - ie_dZ^{`\mu}\phi)^*(\partial_{\mu}\phi - ie_dZ^{`}_{\mu}\phi) - m^2_d\phi^*\phi - \lambda_{\phi}(\phi^+\phi)^2$. 
The annihilation cross section  $\phi_d \bar{\phi_d} \rightarrow \nu_{\mu}\bar{\nu}_{\mu}, \nu_{\tau}\bar{\nu}_{\tau}$  
 for $s \approx 4 m^2_{d}$
has the form\footnote{Here we consider the case $m_{Z^`} > 2 m_{d}$.}
\begin{equation}
\sigma v_{rel} = \frac{8\pi}{3} \frac{\epsilon^2\alpha\alpha_d m^2_{d}v^2_{rel}}
{(m^2_{Z^`} - 4 m^2_{d})^2 } \,,
\end{equation} 
As a consequence of (24) and (42) the estimate for  $\epsilon^2 \alpha_d $ takes the form
\footnote{Here $\epsilon^2 = \frac{\alpha_{\mu}}{\alpha}$, $\alpha_{\mu} = \frac{e^2_{\mu}}{4\pi}$ and
    $\alpha_d = \frac{e_d^2}{4\pi}$.}
\begin{equation}
0.43 \cdot 10^{-6}\cdot(\frac{m_{d}}{GeV})^2 \cdot(\frac{m^2_{Z^`}}{m^2_{d}} - 4)^2 
 = \epsilon^2 \alpha_{d} \,.
\end{equation}
For instance, for $m_{A^`} = 3m_{d}$ we find 
\begin{equation}
1.1  \cdot  10^{-5}\cdot(\frac{m_{d}}{1~GeV})^2 = \epsilon^2 \alpha_{d} \,.
\end{equation}
For $m_{Z^`} \ll m_{\mu}$ 
the values $\epsilon^2 = (2.5 \pm 0.7) \cdot 10^{-6} $ and 
$\alpha_{d} =
(4.4 \pm 1.2)  \cdot (\frac{m_{d}}{1~GeV})^2 $ explain both 
the  $g_{\mu} - 2$ muon anomaly and today dark matter density. 
For $m_{Z^`} = 300~MeV$,  $m_{d} = 100~MeV$ the values $\epsilon^2 =   (4.6 \pm 1.3) \cdot 10^{-5} $,  
$\alpha_{d}= (2.4 \pm 1.1) \cdot 10^{-3} $  lead to correct values for  $g_{\mu}-2$ anomaly and dark matter density.
It should be noted that for  the model with additional scalar dark matter
 the BaBar bound \cite{BABAR1} $m_{Z^`} \leq 212~MeV$ based on the use of  visible decay $Z^` \rightarrow \mu^+\mu^-$ 
in   the reaction $e^+e^- \rightarrow Z'\mu^+ \mu^- $ for the  search 
for $Z'$ boson does not work for $m_{Z`} > 2 m_{d}$  since  
the  $Z^`$ decays mainly into invisible dark matter scalars $Z^` \rightarrow \phi_d \bar{\phi_d} $. The single 
remaining bound 
$m_{Z^`} \leq 400~MeV $ comes from the study of trident muon events in the 
reaction $\nu_{\mu}N \rightarrow \nu_{\mu}N \mu^+\mu^-$ \cite{POSPELOV}.

We can  make an additional assumption that the interaction of the $Z^`$ with dark matter and the $L_{\mu} - L_{\tau}$ 
is universal, i.e. $e_{\mu} = e_d$. This assumption allows to estimate the $m_{Z^`}$ value. We find that for all reasonable 
values of the ratio $\frac{m^2_{Z^`}}{m_d}$ the value of $m_d$ is less than $1~MeV$. For instance, 
for   $\frac{m_{Z^`}}{m_d} = 1.1(2.2)$ we find that $m_{Z^`} = 120~KeV(400~KeV)$.

\section{Conclusions}
We have proposed renormalizable extension of the SM
model  based on the use of additional singlet lepton Higgs isodoublets, scalar isosinglet and dark fermion. 
Light singlet scalar mixes with lepton higgses that induces naturally small 
Yukawa interactions  of scalar singlet  with leptons. 
The induced Yukawa interaction of the isosinglet scalar with muons  can explain  
both $g_{\mu} - 2$ anomaly   without conflict with existing bounds from rare $K$- and $B$-meson  decays. 
Moreover, an additional light isosinglet fermion field 
interacting with singlet scalar  can play a role of 
dark matter and explain today dark matter density.
Also we have mentioned that the addition of  light dark matter scalar in  the model with   vector boson
interacting with $L_{\mu} - L_{\tau}$ current allows to solve the dark matter problem. 
It should be noted that for the models with scalar or vector mediators interacting  mainly with muons
the perspectives to discover dark matter  
in direct underground experiments look very gloomy or even hopeless. One of the possibilities  
to test such models is the use of muon beams \cite{matveev} at 
future  CERN SPS      NA64 experiment \cite{matveev, NA64exp}. Also future BELLE-2    experiment can 
use the reaction $e^+e^- \rightarrow \gamma(Z^` \rightarrow invisible)$ 
to search for $Z^`$ boson \cite{japan}.   

I am indebted to my colleagues from INR Th department for useful discussions.

\section{ Appendix A: Muon $g-2$ contributions}

In this appendix we collect the main formulae for one loop contribution to muon $g-2$ 
anomalous magnetic moment due to existence of new vector(scalar) interactions.
Vector boson (dark photon) which couples very weakly with muon with $\alpha_{Z'} \sim O(10^{-8})$ 
can explain $(g_{\mu}-2)$ anomaly \cite{vecmuon1} - \cite{vecmuon8}.  
Vectorlike interaction of $Z'$ boson with muon  
\begin{equation}
L_{Z'} = g^`_V\bar{\mu}\gamma^{\mu}\mu Z'_{\mu}\,
\end{equation}
leads to  additional contribution to  muon 
anomalous magnetic moment \cite{muonmoment}     
\begin{equation}
\delta a  = \frac{\alpha^`_V}{2\pi} F(\frac{m_{Z'}}{m_{\mu}}) \,,
\end{equation}
where
\begin{equation}
F(x) = \int^1_0 dz \frac{[2z(1-z)^2]}{[(1-z)^2 + x^2z]} \,
\end{equation}
and $\alpha^`_V = \frac{(g^`_V)^2}{4\pi}$. Note that very often data are analyzed in terms of new
variable  $\epsilon^2_V = \frac{\alpha^`_V}{\alpha}$  ($\alpha = 1/137$).
  The use of formulae  (28,29) allows to determine the coupling constant $\alpha^`_V$ which 
explains the value (1) of muon anomaly. For $ m_{Z'} \ll m_{\mu}$  \cite{muonmoment} 
\begin{equation}
\alpha^`_V = (1.8 \pm 0.5) \times 10^{-8}\\.
\end{equation}
For another limiting case  $ m_{Z'} \gg m_{\mu}$ the $\alpha^`_V$ is 
\begin{equation}
\alpha^`_V = (2.7 \pm 0.7) \times 10^{-8} \times \frac{m^2_{Z'}}{m^2_{\mu}}   \\.
\end{equation}
The Yukawa interaction of the scalar field with muon
\begin{equation}
L_{Yuk, \phi} = -g_{\mu \phi} \phi \bar{\mu}\mu  \,.
\end{equation}
leads to  additional one loop contribution to  muon 
anomalous magnetic moment \cite{muonmoment}     
\begin{equation}
\Delta a_{\mu} = \frac{{g}^2_{\mu \phi}}{8\pi^2}\frac{m^2_{\mu}}{m^2_{\phi}}
\int^1_0\frac{x^2(2-x)dx}{(1-x)(1-\lambda^2x) + \lambda^2 x} \,,
\end{equation}
where $\lambda =  \frac{m_{\mu}}{m_{\phi}}$. 
For heavy scalar $m_{\phi} >> m_{\mu}$ 
\begin{equation}
\Delta a_{\mu} = \frac{{g}^2_{\mu \phi}}{4\pi^2}\frac{m^2_{\mu}}{m^2_{\phi}} [ln(\frac{m_{\phi}}{m_{\mu}}) 
- \frac{7}{12}] \,
\end{equation} 
and for light scalar $m_{\mu} \gg m_{\phi}$  
\begin{equation}
\Delta a_{\mu} = \frac{3{g}_{\mu \phi}^2}{16\pi^2}  \,.
\end{equation}

\section{Appendix B: Dark matter density computation}

Here we collect the main formulae for the calculation of today dark matter density. 
To estimate the dark matter density we 
assume  that in the hot early Universe dark matter is in equilibrium with 
ordinary matter. During the Universe expansion the temperature decreases and at some 
temperature $T_d$ the thermal decoupling of the dark matter starts to work \cite{Universe1} -\cite{Universe3}. Namely, at 
some freeze-out temperature the annihilation cross-section 
$\sigma(DMparticles  ~~\rightarrow 
~~SM particles)~$  becomes too small to obey the equilibrium of dark matter particles with 
the SM particles and dark matter decouples. 
To  obtain quantitative estimates of  the  dark matter density \cite{Universe1} -\cite{Universe3} 
 it is necessary  to solve the Boltzmann equation 
\begin{equation}
\frac{dn_{d}}{dt} + 3H(T)n_{d} = - <\sigma v_{rel}>(n^2_{d} - n^2_{d,eq})\,.
\end{equation}
The approximate  solution of the  Boltzmann equation 
can be represented in the form \cite{gondolo}
\begin{equation}
\Omega_{d}h^2 =8.76 \times 10^{-11} GeV^{-2}[ \int^{T_{d}}_{T_0}( g^{1/2}_* <\sigma v_{rel}>)\frac{dT}{m_{d}} ]^{-1}
\end{equation}
The ratio  of dark particle mass $m_{d}$ and freeze-out temperature $T_{d}$  depends logarithmically 
on the $m_{d}, ~< \sigma v_{rel}>, ~T_{d} $, namely \cite{dr}
\begin{equation}
\frac{m_{d}}{T_{d}} \approx 17 + ln(<\sigma v_{rel}>/10^{-26}cm^3~s^{-1}) +
ln(m_{d}/GeV) + ln(\sqrt{m_{d}/T_{d}}) \,.
\end{equation}
The formulae (37,38) for dark matter density and for $\frac{m_{d}}{T_{d}}$  allow to estimate the annihilation cross section.
Numerically for $\frac{m_{d}}{T_{d}} \approx 14$ and  $g^{1/2}_* = 3.7$
and $ \Omega_{d} h^2 = 0.12$ \cite{partdata} the estimate\footnote{
The estimate  $g^{1/2}_* = 3.7$ is valid  for $5~MeV < m_{d} < 2000~ MeV $. For $m_{d} < 5~MeV $ 
electrons don't give significant contribution into $g^{1/2}_*$ and  $g^{1/2}_* \approx 2.7 $.}
for the $s$-wave cross section ($\sigma v_{rel} = const $) is 
\begin{equation}
<\sigma v_{rel}> =  0.28 \cdot 10^{-8}GeV^{-2}
\end{equation}
For p-wave cross section $\sigma v_{rel}^2 = B v^2$ 
\begin{equation}
B = <\sigma v_{rel}/><v^2_{rel}> =  1.3 \cdot 10^{-8}GeV^{-2}
\end{equation}

For the case where dark matter consists of dark matter particles and dark matter antiparticles  $\sigma = \frac{\sigma_{an}}{2}$, 
where $\sigma_{an}$  is the $DM \bar{DM} \rightarrow SM~ particles$ annihilation cross sestion.
For numerical estimates we 
use the values 
\begin{equation}
<\sigma_{an} v_{rel}>  = 0.56\cdot 10^{-8}~GeV^{-2} \,,
\end{equation}
for the $s$-wave cross section and
\begin{equation}
B_{an} = <\sigma_{an} v_{rel} >/<v^2_{rel}> =  2.6 \cdot 10^{-8}~GeV^{-2} \,
\end{equation}
for $p$-wave cross section.

\end{document}